\documentclass[3p,times]{elsarticle}

\usepackage{ecrc}
\usepackage{setspace}
\volume{00}

\firstpage{1}

\runauth{}

\jid{procs}

\usepackage{amssymb}

\usepackage{lineno}
\usepackage{amsmath}
\usepackage{bm}
\usepackage{color}

\usepackage[figuresright]{rotating}

\newcommand{\red}{}
\newcommand{\redd}{}

\begin{document}

\begin{frontmatter}



\dochead{}

\title{Imaging reconstruction method on X-ray data of CMOS polarimeter combined with coded aperture}


\author[isas,utokyo]{Tsubasa Tamba}
\author[utokyo,osakau,ipmu]{Hirokazu Odaka}
\author[utokyo]{Taihei Watanabe}
\author[utokyo]{Toshiya Iwata}
\author[utokyo]{Tomoaki Kasuga}
\author[utokyo,kagoshimau]{Atsushi Tanimoto}
\author[utokyo]{Satoshi Takashima}
\author[utokyo]{Masahiro Ichihashi}
\author[isas]{Hiromasa Suzuki}
\author[utokyo,resceu,tsqsi]{Aya Bamba}

\address[isas]{Japan Aerospace Exploration Agency, Institute of Space and Astronautical Science, 3-1-1, Yoshino-dai, Chuo-ku, Sagamihara, Kanagawa 252-5210, Japan}
\address[utokyo]{Department of Physics, Faculty of Science, The University of Tokyo, 7-3-1, Hongo, Bunkyo-ku, Tokyo 113-0033, Japan}
\address[osakau]{Department of Earth and Space Science, Osaka University, 1-1 Machikaneyama-cho, Toyonaka, Osaka 560-0043, Japan}
\address[ipmu]{Kavli IPMU, The University of Tokyo, Kashiwa 113-0033, Japan}
\address[kagoshimau]{Graduate School of Science and Engineering, Kagoshima University, 1-21-24, Korimoto, Kagoshima, Kagoshima 890-0065, Japan}
\address[resceu]{Research Center for Early Universe, Faculty of Science, The University of Tokyo, 7-3-1, Hongo, Bunkyo-ku, Tokyo 113-0033, Japan}
\address[tsqsi]{Trans-Scale Quantum Science Institute, The University of Tokyo, Tokyo  113-0033, Japan}

\begin{abstract}

X-ray polarization is a powerful tool for unveiling the anisotropic characteristics of high-energy celestial objects.
We present a novel imaging reconstruction method designed for hard X-ray polarimeters employing a Si CMOS sensor and coded apertures, which function as a photoelectron tracker and imaging optics, respectively.
Faced with challenges posed by substantial artifacts and background noise in the coded aperture imaging associated with the conventional balanced correlation method, we adopt the Expectation-Maximization (EM) algorithm as the foundation of our imaging reconstruction method.
\red{The newly developed imaging reconstruction method is validated with imaging polarimetry and a series of X-ray beam experiments.}
The method demonstrates the capability to accurately reproduce an extended source comprising multiple segments with distinct polarization degrees.
Comparative analysis exhibits a significant enhancement in imaging reconstruction accuracy compared to the balanced correlation method, with the background noise levels reduced to 17\%.
\red{The outcomes of this study enhance the feasibility of Cube-Sat imaging polarimetry missions in the hard X-ray band, as the combination of Si CMOS sensors and coded apertures is a promising approach for realizing it.}


\end{abstract}


\end{frontmatter}



\section{Introduction}\label{sec:intro}

X-ray polarization is a distinctive method for capturing unprecedented images of high-energy celestial objects.
It carries essential information about anisotropic physical phenomena, such as the magnetic field structure responsible for synchrotron radiation, the emission geometry of Compton scattering, and the properties of the strong gravitational field around compact objects.
Recent advancements in X-ray polarization studies have been marked by the launch of the Imaging X-ray Polarimeter Explorer (IXPE; \citep{Weisskopf2016, Weisskopf2022}) in 2021.
This mission, which covers the soft X-ray band of 2--8 keV, has substantially broadened our capabilities in this field (e.g., Crab Nebula observation: \citep{Bucciantini2023}).
While IXPE focuses on the soft X-ray band through photoabsorption, the Soft Gamma-ray Detector \red{(SGD)} onboard Hitomi \citep{Tajima2018} and the Polarized Gamma-ray Observer+ (PoGO+;\citep{Chauvin2017_mission}) utilize Compton scattering to detect X-ray polarization in the hard X-ray band.
Notably, both missions successfully detected polarization from the Crab Nebula in the higher end of the hard X-ray band: 60--160 keV for SGD \citep{Hitomi2018} and 20--160 keV for PoGO+ \citep{Chauvin2017_crab}.
However, a significant observational gap exists in the lower end of the hard X-ray band due to the absence of established observational techniques, specifically in the energy range of 10--30 keV.
This energy range is particularly important because non-thermal components, carrying anisotropic information, become dominant over unpolarized thermal emission above $\sim10\;{\rm keV}$.
In addition, the abundance of photon counts in this band, compared to higher energy bands, provides a rich dataset for detailed analysis.

To address this observational gap, \red{a 6U CubeSat mission named the Coded Imaging Polarimeter of High Energy Radiation (cipher; \citep{Odaka2020}) is under development}.
It aims to conduct imaging polarimetry in the 10--30 keV band and obtain the polarization map for bright extended sources such as the Crab Nebula.
This imaging polarimeter is characterized by a micro-pixel \red{($\sim{\rm \mu m}$ pixel pitch) Si} CMOS sensor and coded apertures.
The micro-pixel CMOS sensor serves as a sophisticated photoelectron tracker, enabling the determination of the polarization angle of incident photons.
Given the spatial constraints of the CubeSat mission, coded apertures provide the solution for imaging capability, eliminating the need for X-ray mirrors.
The CMOS sensor has demonstrated polarization detection capabilities in the 6--30 keV range \citep{Asakura2019, Iwata2024}, meeting the fundamental requirements of cipher.
The imaging reconstruction of the coded apertures is also established \citep{Kasuga2020}, employing a conventional reconstruction technique called the ``balanced correlation method'' \citep{Fenimore1978}.
However, this method introduces significant artifacts and noise levels, which hinder effective imaging polarimetry.

Recognizing the limitation of the balanced correlation method, this paper presents a novel imaging reconstruction method for polarimetry utilizing the Expectation-Maximization (EM) algorithm \citep{Dempster1977}.
It is a statistical approach to efficiently derive maximum likelihood estimates, and has been applied to imaging reconstruction techniques \red{extensively} (e.g., \citep{Ikeda2014, Reader2002, Fessler1995, Maher1985, Zanetti2015}).
This method holds the potential to achieve imaging polarimetry with enhanced precision, \red{with} reduced artifacts and noise levels.
The remainder of this paper is organized as follows.
We first provide the formulation of the EM algorithm \red{that is applied} to imaging polarimetry, and describe \red{the experimental} setup to examine the new method in Section \ref{sec:method}.
The data processing and results of imaging reconstruction are presented in Section \ref{sec:results}.
We then discuss the precision of our new imaging reconstruction method by comparing it with the conventional balanced correlation method in Section \ref{sec:discussion}.
Conclusions are presented in Section \ref{sec:conclusion}.




\section{Imaging reconstruction method}\label{sec:method}


\subsection{Application of EM algorithm to imaging polarimetry}

We developed a novel reconstruction method for the coded aperture imaging based on the \red{EM} algorithm \citep{Dempster1977}.
This statistical approach involves estimating the maximum likelihood from incomplete experimental data through iterative cycles of Expectation (E-) and Maximization (M-) steps, continuously updating the estimation.
The $l$-th E-step and M-step in the imaging reconstruction analysis are defined as follows:
\begin{eqnarray}
{\rm (E-step)}\;\;\;\;\;\tilde{D}_v^{(l)}&=&\sum_{u}p(v\,|\,u)\tilde{S}_u^{(l)},\label{eq:estep1}\\
{\rm (M-step)}\;\;\;\;\;\tilde{S}_u^{(l+1)}&=&\frac{1}{\sum_{v'}D_{v'}}\sum_vD_v\frac{p(v\,|\,u)\tilde{S}_u^{(l)}}{\tilde{D}_v^{(l)}}\label{eq:mstep1},
\end{eqnarray}
where $l$ increments by 1 at each iteration \citep{Ikeda2014}.
In these equations, $\tilde{S}_u^{(l)}$, $\tilde{D_v^{(l)}}$, and $D_{v}$ represent the estimated incident source distribution, the estimated event distribution on the detector, and the actual event distribution derived from experimental data, respectively.
The term $p(v\,|\,u)$ denotes the posterior probability that a photon from the $u$-th sky region is detected in the $v$-th detector region.

\red{The method is applied to the imaging polarimeter with a CMOS sensor and coded apertures.}
The sky region is defined as a two-dimensional angular distribution, $\bm{u}=(\theta_x,\;\theta_y)$, with respect to the optical axis, while the detector region is defined as a two-dimensional pixel distribution, $\bm{v}=(d_x,\;d_y)$.
The polarization angle and polarization degree need to be assigned to each sky region.
Since the CMOS sensor is sensitive only to the horizontal and vertical directions, we focused on two Stokes parameters, $I$ and $Q$, ignoring the effect of $U$.
These Stokes parameters are linked to the experimental data through the branching ratio in double-pixel events, \red{which is}, the ratio between horizontally-elongated (H-type) and vertically-elongated (V-type) events \citep{Odaka2020}.
Equations \eqref{eq:estep1} and \eqref{eq:mstep1} can be expressed as
\begin{eqnarray}
\tilde{D}_{\bm{v},\,t}^{(l)}&=&\sum_{\bm{u}}p(\bm{v}\,|\,\bm{u})\tilde{S}_{\bm{u},\,t}^{(l)}\label{eq:estep2}\\
\tilde{S}_{\bm{u},\,t}^{(l+1)}&=&\frac{1}{\sum_{\bm{v}'}D_{\bm{v}',\,t}}\sum_{\bm{v}}D_{\bm{v},\,t}\frac{p(\bm{v}\,|\,\bm{u})\tilde{S}_{\bm{u},\,t}^{(l)}}{\tilde{D}_{\bm{v},\,t}^{(l)}},\label{eq:mstep2}
\end{eqnarray}
where $t$ is a two-valued variable representing $t={\rm H-type}$ or $t={\rm V-type}$.
The Stokes parameters $I$ and $Q$ can be calculated using
\begin{eqnarray}
\begin{pmatrix}
I_{\bm{u}}^{(l)} \\
Q_{\bm{u}}^{(l)} \\
\end{pmatrix}
=
\begin{pmatrix}
1 & 1 \\
1/m & -1/m\\
\end{pmatrix}
\begin{pmatrix}
S_{\bm{u},\,{\rm H-type}}^{(l)}\\
S_{\bm{u},\,{\rm V-type}}^{(l)}\\
\end{pmatrix}
,
\end{eqnarray}
where $m$ denotes the modulation factor of the CMOS sensor.
The posterior probability $p(\bm{v}\,|\,\bm{u})$ is calculated from the coded aperture pattern:
\begin{eqnarray}
p(\bm{v}\,|\,\bm{u})\propto
\begin{cases}
1 & (\bm{u}\;{\rm to}\;\bm{v}\;{\rm pass\;through\;aperture})\\
\tau & (\bm{u}\;{\rm to}\;\bm{v}\;{\rm pass\;through\;mask})\\
0 & {\rm (otherwise)},
\end{cases}
\end{eqnarray}
where $\tau$ is the transmittance of the coded aperture.


\subsection{Beam experiment}

\begin{figure*}[htbp]
\centering
\includegraphics[width=\linewidth]{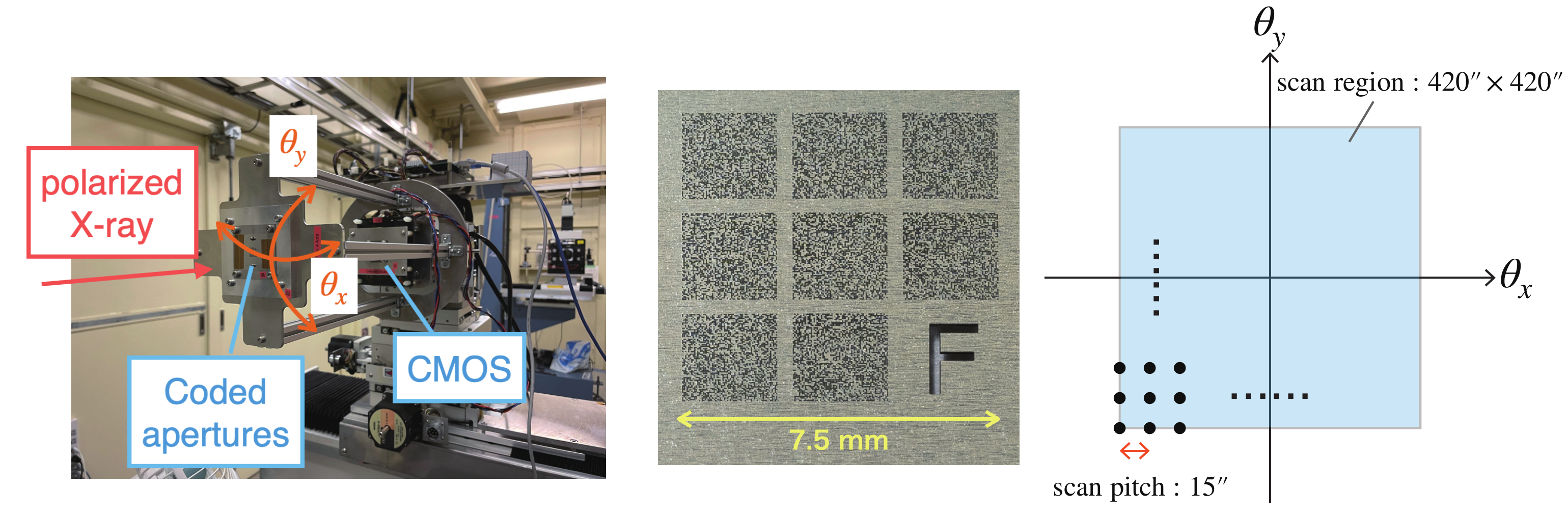}
\caption{
(left) Imaging polarimetry setup at Spring-8 beam \red{line}.
(middle) Picture of coded apertures.
(right) Sky region subject to the imaging scan experiment.
}
\label{fig:experiment_setup}
\end{figure*}

\red{X-ray beam experiments were conducted} at BL20B2 in SPring-8, a synchrotron radiation facility in Hyogo, Japan, to evaluate the imaging-polarimetry capability of \red{the} system.
The beamline emits polarized monochromatic X-rays with a polarization degree of nearly 100\% (e.g., \citep{Asakura2019}).
The left panel of Figure \ref{fig:experiment_setup} shows \red{the} experimental setup at the beamline.
\red{We utilized the GMAX0505RF sensor, manufactured by GPixel Inc., as the polarization detector.
This Si CMOS sensor, with a pixel pitch of $2.5\;{\rm \mu m}$, was originally designed for visible light and near-infrared imaging but also has sensitivity for the X-ray polarization \citep{Iwata2024}.
The sensor was positioned at the farthest part from the beam source.
Upstream of it, a $25\;{\rm cm}$ collimator was mounted to collimate incident X-rays.
On the upstream side of the collimator,} a $0.1\;{\rm mm}$ thick board made of SUS304 was attached.
This board features numerous holes, functioning as eight independent coded apertures with distinct random patterns.
The collimator ensures that the shadows of these patterns do not overlap at the detector plane.
Each coded aperture comprises \red{of} a $64\times64$ pattern with a pitch \red{size} of $35\;{\rm \mu m}$, covering $896\times896$ pixels on the detector plane (Figure \ref{fig:experiment_setup} middle).
The system has a field of view and angular resolution of $0.5^{\circ}\times0.5^{\circ}$ and $29''$, respectively.
The orientation of the collimator ($\theta_x$ and $\theta_y$) is adjustable using a rotation stage, facilitating the control of the incident X-ray direction.
Moreover, the entire system can rotate around the beam axis, allowing for the \red{adjustment} of the polarization angle detected by the CMOS sensor.

\red{A comprehensive sky scan was conducted by manipulating the rotation stage, and an image for an extended source was created by merging the acquired data.}
The data points for \red{the} imaging scan experiments are illustrated in the right panel of Figure \ref{fig:experiment_setup}, covering a $420''\times420''$ plane with a pitch \red{size} of $15''$.
A total of $29\times29=841$ \red{scan points} were utilized.
Table \ref{tab:data_list} provides \red{the} details on the datasets employed in this paper.
To verify the hard X-ray imaging polarimetry, we exposed the system to $16.0\;{\rm keV}$ polarized X-rays with polarization angles of both $0^{\circ}$ and $90^{\circ}$.
\red{This beam energy was selected by balancing between the polarization detection capability and the quantum efficiency of the sensor.
Lower-energy photons would result in low sensitivity for polarization detection because they are more likely to be recorded as single-pixel events.
Conversely, higher-energy photons lead to poor statistical quantity due to their smaller cross section with the Si CMOS sensor.}
The incident beam was attenuated by a $120\;{\rm \mu m}$ Cu filter to avoid pile-up.
Each dataset \red{has} 10 frames per scan point, resulting in a total of 8410 frames.
The initial four datasets in Table \ref{tab:data_list} were merged to generate the ``$0^{\circ}$ polarization data'', while the latter four were merged to generate the ``$90^{\circ}$ polarization data''.


\begin{table*}[htbp]
\centering
\fontsize{11}{13}\selectfont
\caption{List of \red{datasets}.
}
\begin{tabular}{ccccc}
\hline \hline
\begin{tabular}{c} Start time\\(JST)\\\end{tabular} & \begin{tabular}{c}Beam energy\\(keV)\\ \end{tabular} & \begin{tabular}{c}Polarization angle\\(degree)\end{tabular} & \begin{tabular}{c}Frame exposure\\(ms)\\\end{tabular} & Number of frames\\
\hline
2021-11-03 18:50 & 16.0 & 0 & 600 & 8410\\
2021-11-03 20:40 & 16.0 & 0 & 600 & 8410\\
2021-11-03 22:42 & 16.0 & 0 & 600 & 8410\\
2021-11-04 00:22 & 16.0 & 0 & 600 & 8410\\
2021-11-04 03:42 & 16.0 & 90 & 600 & 8410\\
2021-11-04 05:23 & 16.0 & 90 & 600 & 8410\\
2021-11-04 07:00 & 16.0 & 90 & 600 & 8410\\
2021-11-04 08:54 & 16.0 & 90 & 600 & 8410\\
\hline
\end{tabular}
\label{tab:data_list}
\end{table*}

\section{Results}\label{sec:results}

\subsection{Data processing}\label{subsec:data_processing}

We performed standard data processing on the acquired data using ComptonSoft \citep{Odaka2010, Suzuki2020, Tamba2022}, which included pedestal subtraction, bad pixel exclusion, event extraction, and event classification.
Subsequent to \red{the} data processing, \red{double-pixel events exhibiting either horizontal (H-type) or vertical (V-type) elongation were specifically selected}.
The polarization angle and polarization degree can be determined by measuring the ratio between \red{the} two types of events.
\red{The numbers of H-type/V-type events were 9485530/8008517 for $0^{\circ}$ polarization and 7622987/9741000 for $90^{\circ}$ polarization.}
\red{The modulation factor was calculated from these values to be} $m=0.1033\pm0.0001$ for $16.0\;{\rm keV}$ polarized photons (for more details of the modulation factor, see \citep{Iwata2024}).

\begin{figure}[htbp]
\centering
\includegraphics[width=210.0pt]{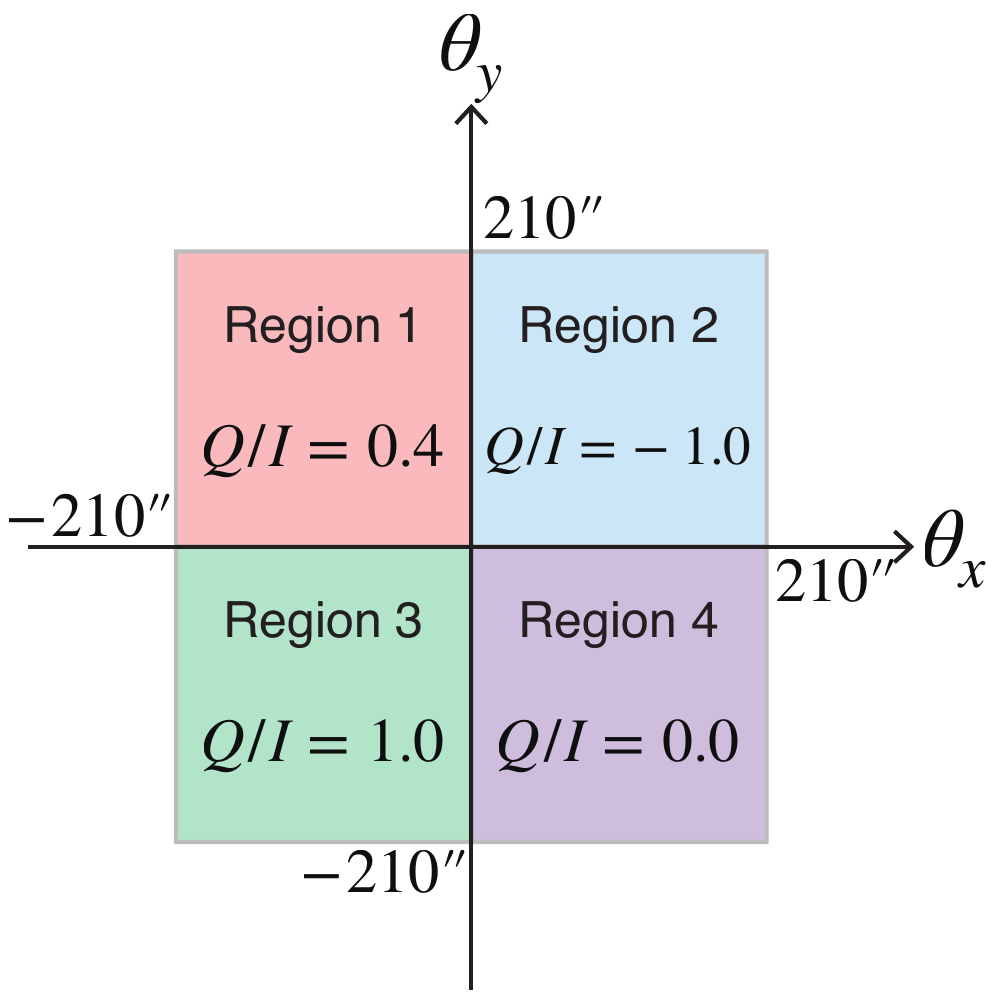}
\caption{Definition of the imaging polarimetry test data (see Section \ref{subsec:data_processing}).
}
\label{fig:region_definition}
\end{figure}

In addition to the $0^{\circ}$ and $90^{\circ}$ polarization data, a new dataset \red{named} ``imaging polarimetry test data'' \red{with} various polarization degrees depending on incident directions \red{was generated}.
This dataset \red{can easily be} generated by appropriately blending the $0^{\circ}$ and $90^{\circ}$ polarization data; for instance, an unpolarized source can be simulated by mixing equal proportions of the $0^{\circ}$ and $90^{\circ}$ polarization data.
To evaluate the imaging polarimetry capability, we divided the entire scan region into four sub-regions and assigned distinct polarization degrees ($Q/I$) to \red{them}.
Figure \ref{fig:region_definition} illustrates the definition of the segmentation.
We assigned $Q/I=0.4$, $-1.0$ (\red{$90^{\circ}$} polarization), $1.0$ (\red{$0^{\circ}$} polarization), and $0.0$ (unpolarized) to Regions 1, 2, 3, and 4, respectively, where the intensity $I$ was spatially uniform.
\red{The first value ($Q/I=0.4$) was assigned to examine a ``realistic'' polarization degree for a celestial object, while the latter three values ($Q/I=-1.0$, $1.0$, $0.0$) were included to examine extreme cases.}

\subsection{Imaging polarimetry}

\red{The EM algorithm was applied} to the reconstruction of the polarization map from the encoded images on the detector plane.
We set $\tilde{S}_{\bm{u},\,t}^{(0)}$ as a spatially uniform distribution, and iterated through the expectation and maximization steps using equations \eqref{eq:estep2} and \eqref{eq:mstep2}.
The reconstruction was simultaneously applied to all eight coded apertures, with the transmittance of the apertures set to $\tau=0.02$ (corresponding to $0.1\;{\rm mm}$ SUS304 for $16\;{\rm keV}$ photons).
The encoded image on the detector plane was binned \red{in} every $4\times4$ pixels due to the limited memory resources, but \red{this} did not affect the results as the angular resolution is primarily dominated by the aperture pitch.
\red{The reconstructed images were generated with an image pixel size of $20''\times20''$, which is slightly smaller than the angular resolution of the system.}
The EM steps were iterated until $l=1500$, ensuring a sufficient convergence.
The reconstructed images were obtained after the vignetting correction was applied.

\begin{figure*}[htbp]
\centering
\includegraphics[width=\linewidth]{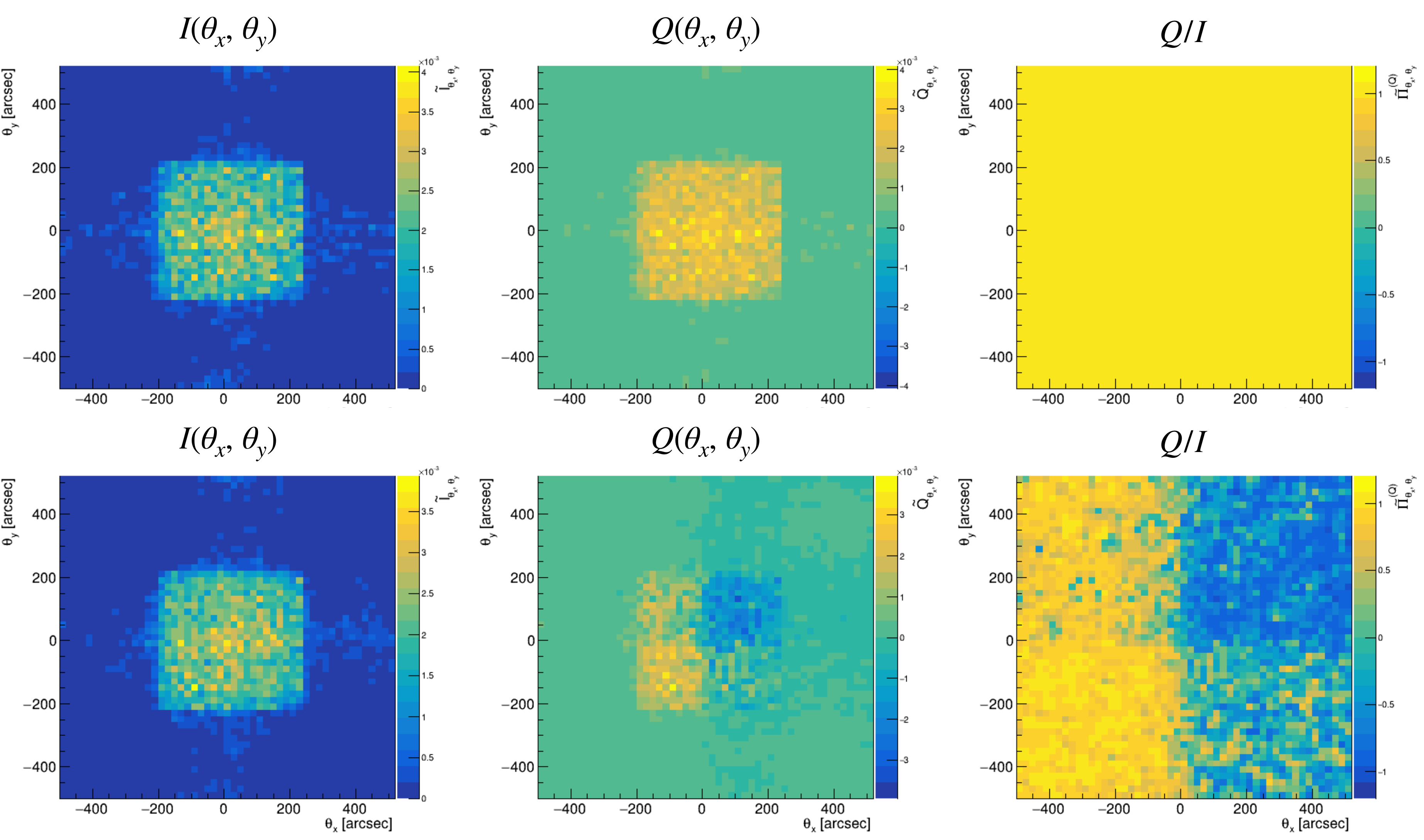}
\caption{The upper panels show the reconstructed images from $0^{\circ}$ polarization data, while the lower panels show those from imaging polarimetry test data (see text).
Reconstructed images include $I(\theta_x,\;\theta_y)$ (left), $Q(\theta_x,\;\theta_y)$ (middle), and polarization degree $Q/I$ (right).
}
\label{fig:EM_result}
\end{figure*}

As a simple case, we first conducted the imaging reconstruction on the $0^{\circ}$ polarization data, where $I$ and $Q$
distributions should ideally be identical.
The upper panels of Figure \ref{fig:EM_result} display the reconstructed images of $I$, $Q$, and $Q/I$.
\red{The uniformly distributed images of $I$ and $Q$ were successfully reconstructed}, consistent with the actual scan region of $420''\times420''$.
The polarization degrees also yield a convincing value of $Q/I=1$ throughout the entire image.
It is important to note that the $Q/I$ image indicates horizontal polarization of incident photons even beyond the scan region.
This artifact arises because a portion of \red{the} detected photons are inaccurately projected to the outer region.
Consequently, careful examination of both $I$ and $Q/I$ images is necessary, and polarization degrees should not be relied on in regions where the photon intensity $I$ is faint.
\red{We also tried various patterns of uniformly distributed $Q/I$ values (0.5, 0, -0.5, -1.0) and confirmed that the reconstructed images successfully reproduced the uniformly distributed polarization degrees.}

Subsequently, we applied the imaging reconstruction to the imaging polarimetry test data, which is described in Section \ref{subsec:data_processing} and defined in Figure \ref{fig:region_definition}.
The lower panels of Figure \ref{fig:EM_result} display the reconstructed images.
\red{A} uniform distribution of $I$ extending over $420''\times420''$ similar to the $0^{\circ}$ polarization data \red{was observed}.
\red{But a totally different image of $Q$, clearly divided into four segments, was observed.}
Each of the four regions corresponds to the definition in Figure \ref{fig:region_definition}.
The reconstructed image of $Q/I$ successfully reflects the distinct polarization degrees of the four regions, displaying $\sim0.4$, $\sim-1.0$, $\sim1.0$, and $\sim0.0$ for Regions 1, 2, 3, and 4, respectively.

\begin{figure*}[htbp]
\centering
\includegraphics[width=\linewidth]{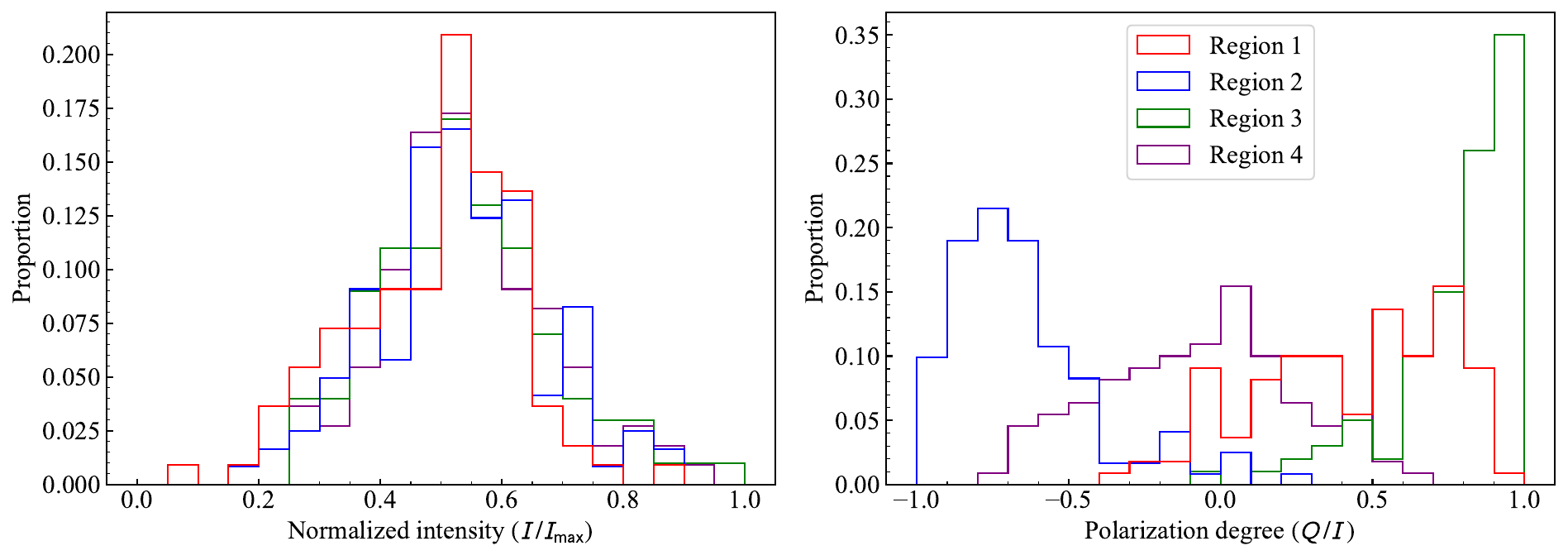}
\caption{Evaluation of imaging reconstruction on the imaging polarimetry test data.
The left panel shows the histogram of source intensity plotted for the Regions 1--4.
The right panel displays the histogram of polarization degree plotted for Regions 1--4.
}
\label{fig:EM_evaluation}
\end{figure*}

In Figure \ref{fig:EM_evaluation}, a detailed evaluation of the imaging reconstruction results \red{is presented}.
The left panel shows the distribution of $I/I_{\rm max}$ for the four regions, where $I_{\rm max}$ corresponds to the maximum value of the entire image.
The four regions display comparable distributions, indicating that the spatially uniform distribution of $I$ is successfully reproduced.
The right panel exhibits the distribution of the polarization degree $Q/I$ for the four regions.
The four regions display peaks at the predetermined polarization degrees and are significantly distinguishable from each other.
\red{Still, non-negligible uncertainties of $\Delta(Q/I)\sim0.2$ remain due to systematic uncertainties stemming from potentially non-uniform modulation factors across the detector plane and biased projections caused by randomized patterns of the coded apertures.
While a poorer statistical quantity would result in worse reproducibility, this imaging analysis benefits from an average of $\sim40000$ photon counts contributing each imaging pixel.}

\section{Discussion}\label{sec:discussion}

In Section \ref{sec:results}, \redd{our formulation of the imaging reconstruction method using the EM algorithm was examined, and the polarization map was successfully reconstructed from the imaging polarimetry test data}.
In this section, the accuracy of the imaging reconstruction \red{was evaluated} by comparing our method with the conventional balanced correlation method \citep{Fenimore1978}.
The balanced correlation method is a prominent decoding method for coded apertures, especially suitable for Uniformly Redundant Array (URA; \citep{Fenimore1981, Gottesman1989}), but it also works for random pattern arrays used in our system (for details of the balanced correlation method in our system, see Section 2 of \citep{Kasuga2020}).

\begin{figure*}[htbp]
\centering
\includegraphics[width=\linewidth]{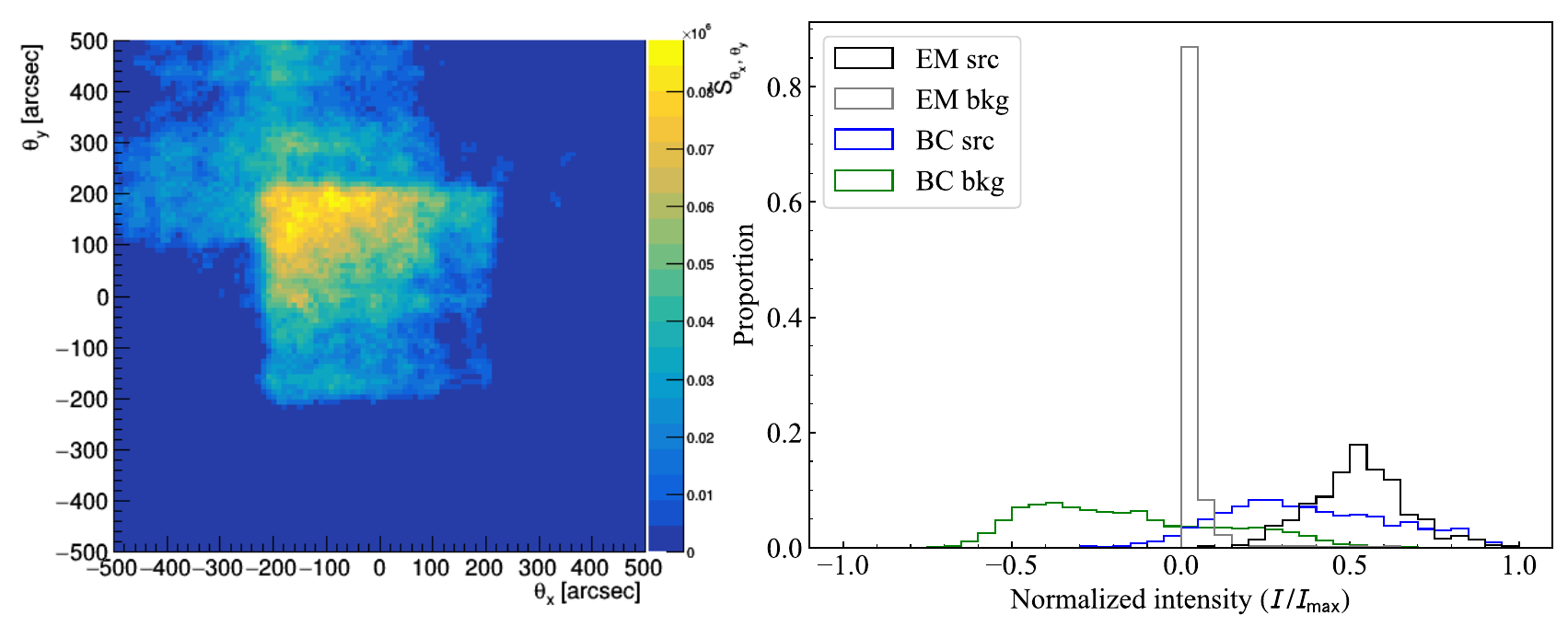}
\caption{
(left) Reconstructed $I$ image of $0^{\circ}$ polarization data using the balanced correlation method.
(right) Intensity distributions of source (src) and background (bkg) regions compared between the EM algorithm (labeled as EM) and the balanced correlation method (labeled as BC).
}
\label{fig:discussion1}
\end{figure*}

The left panel of Figure \ref{fig:discussion1} shows the reconstructed $I$ image of the $0^{\circ}$ polarization data using the balanced correlation method.
Despite the incident flux being spatially uniform along the scan region, it exhibits a significant imbalance with unpredictably intense signals in the upper left region.
The non-source region also displays large fluctuations with a noisy background in the upper left region.
These artifacts and noisy background signals arise from the spatially non-uniform distribution of the apertures, which could back-project more signals to the upper left region of the sky.
In the right panel of Figure \ref{fig:discussion1}, the intensity distributions of the source and background regions \red{are presented}, comparing the results between the EM algorithm and the balanced correlation method.
The EM algorithm exhibits sharp distributions in both source and background regions, with the two distributions clearly separated.
On the other hand, the balanced correlation method has flat distributions for the source and background regions, and these two distributions are largely overlapped.
This difference indicates fewer artifacts and lower noise levels in the EM algorithm compared to the balanced correlation method.

The noise levels can be evaluated by 
\begin{eqnarray}
L_{\rm noise}=\frac{\sigma_{\rm bkg}}{\mu_{\rm src}},
\end{eqnarray}
where $\sigma_{\rm bkg}$ denotes the standard deviation in the background region, and $\mu_{\rm src}$ is the averaged intensity in the source region.
\red{Here, the source region represents a collection of the image pixels within the central $420''\times420''$ area while the background region denotes a collection of the image pixels outside this area.}
We obtain $L_{\rm noise}=0.13$ for the EM algorithm and $L_{\rm noise}=0.75$ for the balanced correlation method, which means that our new imaging reconstruction method reduces the noise level to $17\%$ of the conventional method.
The significant artifacts and noise levels in the balanced correlation method had prevented us from creating any efficient polarization map, but the improved method using the EM algorithm has successfully realized it with sufficiently reduced artifacts and noise levels, as depicted in Figures \ref{fig:EM_result} and \ref{fig:EM_evaluation}.

\red{
Several aspects still require refinement for the application of this system and analysis method to imaging polarimetry.
Firstly, while we assumed a uniform modulation factor within the CMOS sensor, this assumption needs careful examination.
Deviations of $\sim0.01$ from the adopted original value ($m=0.1033$) could occur when focusing on specific limited areas of the sensor.
A better understanding of the dependence of the modulation factor on the sensor pixels would mitigate systematic uncertainties in imagin polarimetry.
Secondly, the biased random patterns of the coded apertures also require scrutiny.
As observed in the left panel of Figure \ref{fig:discussion1}, generated by the balanced correlation method, these coded apertures tend to project more photons to the upper left region of the reconstructed image, potentiallly imacting the accuracy of the imaging reconstruction in the case of the EM algorithm.
It is still underway to develop more sophisticated coded aperture patterns.
Furthermore, the performance of imaging polarimetry is influenced by the shape of the polarization distribution.
While in this study, the boundaries between different polarization profiles are located around the center of the field of view, we also confirmed that the reconstruction accuracy diminishes when these boundaries are located farther from the center.
This is due to reduced sensitivity of the coded apertures in these regions.
These considerations are beyond the scope of this paper and will be addressed in future discussions.
}



\section{Conclusions}\label{sec:conclusion}

We developed a new imaging reconstruction method for hard X-ray imaging polarimetry employing a combination of a CMOS sensor and coded apertures.
Motivated by the significant artifacts and background noise levels associated with the conventional balanced correlation method, we introduced the \red{EM} algorithm as the foundation of our new imaging reconstruction method.
The effectiveness of the newly developed method was confirmed through X-ray beam experiments at Spring-8, where the imaging polarimeter was exposed to the X-ray beam and captured an extended source by a comprehensive scan of the sky.
The method exhibited remarkable capabilities in accurately reproducing an extended source comprising multiple segments characterized by distinct polarization degrees.
Specifically, it successfully reproduced four regions with $Q/I=-1.0$, 0.0, 0.4, 1.0.
Our developed imaging reconstruction method achieved a significant reduction in artifacts and noise levels compared to the balanced correlation method.
The background noise level experienced a significant reduction to $17\%$.
\red{The outcomes of this study demonstrate sufficient feasibility for the hard X-ray imaging polarimetry utilizing the combination of CMOS sensors and coded apertures.
This approach represents one of the most promising ways to achieve CubeSat missions on hard X-ray polarimetry.}

\section*{Acknowledgment}
We thank the anonymous referees, who improved this work with their valuable comments.
We appreciate the helpful technical support by M.Hoshino and K.Uesugi at SPring-8.
This research is supported by the Japan Society for the Promotion of Science (JSPS) KAKENHI Grant No. 	20J20050, 23KJ2214 (TT), 18H05861, 19H01906, 19H05185, 22H00128, 22K18277 (HO), 20J00119 (AT), and 23KJ0780 (TI).
The synchrotron radiation experiments were performed at BL20B2 in SPring-8 with the approval of the Japan Synchrotron Radiation Research Institute (JASRI) (Proposal No. 2019B1369, 2020A1343, 2021B1542, and 2022B1477).
This research is also supported by Society for Promotion of Space Science (TT).





\bibliographystyle{elsarticle-num}
\bibliography{reference}







\end{document}